\documentclass[12pt]{iopart}
\usepackage{epsfig}
\usepackage{amssymb}

        \newdimen\mysep                
        \newdimen\hmysep
\begin{document}

\newcommand{\BS}{\bigskip}
\def    \be             {\begin{equation}}
\def    \ee             {\end{equation}}
\def    \beq             {\begin{equation}}
\def    \eeq             {\end{equation}}
\def    \ba             {\begin{eqnarray}}
\def    \ea             {\end{eqnarray}}
\def    \beqn           {\begin{eqnarray}}
\def    \eeqn           {\end{eqnarray}}
\def    \beeq           {\begin{eqnarray}}
\def    \eeeq           {\end{eqnarray}}
\def    \nn             {\nonumber}
\def    \=              {\;=\;}
\def    \frac           #1#2{{#1 \over #2}}
\def    \ret            {\\[\eqskip]}
\def    \ie             {{\em i.e.\/} }
\def    \eg             {{\em e.g.\/} }
\def \lsim{\mathrel{\vcenter
     {\hbox{$<$}\nointerlineskip\hbox{$\sim$}}}}
\def \gtrsim{\mathrel{\vcenter
     {\hbox{$>$}\nointerlineskip\hbox{$\sim$}}}}
\def    \bentarrow      {\:\raisebox{1.1ex}{\rlap{$\vert$}}\!\rightarrow}
\def    \rd             {{\mathrm d}}    
\def    \Im             {{\mathrm{Im}}}  
\def    \bra#1          {\mbox{$\langle #1 |$}}
\def    \ket#1          {\mbox{$| #1 \rangle$}}

\def    \kev            {\mbox{$\mathrm{keV}$}}
\def    \mev            {\mbox{$\mathrm{MeV}$}}
\def    \gev            {\mbox{$\mathrm{GeV}$}}


\def    \mq             {\mbox{$m_Q$}}  
\def    \mt             {\mbox{$m_t$}}  
\def    \mb             {\mbox{$m_b$}}  
\def    \mqq            {\mbox{$m_{Q\bar Q}$}}
\def    \mqqsq          {\mbox{$m^2_{Q\bar Q}$}}
\def    \pt             {\mbox{$p_T$}}
\def    \et             {\mbox{$E_T$}}
\def    \xt             {\mbox{$x_T$}}
\def    \xtsq           {\mbox{$x_T^2$}}
\def    \ptsq           {\mbox{$p^2_T$}}
\def    \etsq           {\mbox{$E^2_T$}}
        
\newcommand     \MSB            {\ifmmode {\overline{\rm MS}} \else 
                                 $\overline{\rm MS}$  \fi}
\def    \muf            {\mbox{$\mu_{\rm F}$}}
\def    \mug            {\mbox{$\mu_\gamma$}}
\def    \mufsq          {\mbox{$\mu^2_{\rm F}$}}
\def    \mur            {{\mbox{$\mu_{\rm R}$}}}
\def    \mursq          {\mbox{$\mu^2_{\rm R}$}}
\def    \mul            {{\mu_\Lambda}}
\def    \mulsq          {\mbox{$\mu^2_\Lambda$}}

\def    \bzero          {\mbox{$b_0$}}
\def    \as             {\ifmmode \alpha_s \else $\alpha_s$ \fi}
\def    \asb            {\mbox{$\alpha_s^{(b)}$}}
\def    \assq           {\mbox{$\alpha_s^2$}}
\def \oacube {\mbox{$ {\cal O}(\alpha_s^3)$}}
\def \oaemacube {\mbox{$ {\cal O}(\alpha\alpha_s^3)$}}
\def \oafour {\mbox{$ {\cal O} (\alpha_s^4)$}}
\def \oatwo {\mbox{$ {\cal O} (\alpha_s^2)$}}
\def \oaematwo {\mbox{$ {\cal O}(\alpha \alpha_s^2)$}}
\def \oaemas {\mbox{$ {\cal O}(\alpha \alpha_s)$}} 
\def \oas   {\mbox{$ {\cal O}(\alpha_s)$}}
\def\slash#1{{#1\!\!\!/}}
\def\rt1{\raisebox{-1ex}{\rlap{$\; \rho \to 1 \;\;$}}
\raisebox{.4ex}{$\;\; \;\;\simeq \;\;\;\;$}}
\def\ltap{\raisebox{-.5ex}{\rlap{$\,\sim\,$}} \raisebox{.5ex}{$\,<\,$}}
\def\gtap{\raisebox{-.5ex}{\rlap{$\,\sim\,$}} \raisebox{.5ex}{$\,>\,$}} 
\def\lq{\left[}
\def\rq{\right]}
\def\rg{\right\}}
\def\lg{\left\{}
\def\({\left(}
\def\){\right)}

\newcommand\LambdaQCD{\Lambda_{\scriptscriptstyle \rm QCD}}

\def\naive{na\"{\i}ve}
\def\asp{{\alpha_s}\over{\pi}}
\def\GE{\gamma_E}
\def\half{\frac{1}{2}}
\def\bom#1{\mbox{\bf{#1}}}

\nopagebreak
{\flushright{
        \begin{minipage}{4cm}
         DTP/99/104 \\
        {\tt hep-ph/9912206}\hfill \\
        \end{minipage}        }

}

\title[Results in next-to-leading-log prompt-photon hadroproduction] {Results
in next-to-leading-log prompt-photon hadroproduction~\footnote[1]{Talk given
at the {\sl UK Phenomenology Workshop on Collider Physics}, Durham, UK,
19--24~September 1999. To appear in the Proceeding of the conference. }
}

\author{Carlo Oleari}

\address{Department of Physics, University of Durham, Durham, DH1 3LE,
UK}

\begin{abstract} 
  We present some results on renormalization and factorization scale
  dependence of soft-gluon resummation, at the next-to-leading-logarithmic
  level, in the fixed-target hadro-production cross-section for prompt
  photons.  
\end{abstract}                                                


\section{Introduction}
The phenomenological interest in prompt-photon production in fixed-target
experiments resides mainly in its use as a gluon probe in structure-function
studies.  Prompt-photon production is historically our main source of
information on the gluon parton density at large~$x$ (e.g.  $x>0.2$).
This same region is relevant for hadron colliders in production phenomena at
very large transverse momenta, and thus its understanding is crucial in order
to disentangle possible new physics signals from the QCD background.

In this talk, we consider the effect of soft-gluon resummation in
prompt-photon production~\cite{Laenen98,CMN98} near the ``threshold'' limit,
that is to say for high transverse momentum.  A more extensive treatment of
this subject can be found in Ref.~\cite{photon99}.

\section{Kinematics, notation and NLL resummed cross section}
\label{secnotat}                    
We consider the inclusive production of a single prompt photon in hadron
collisions
\beq
\label{procgamma}
H_1(P_1) + H_2(P_2) \to \gamma(p) + X \,.
\eeq
The colliding hadrons $H_1$ and $H_2$ carry momenta $P_1$ and $P_2$,
and the centre-of-mass energy squared is then given by $S= (P_1+P_2)^2$,
while the photon momentum is $p$.
If $\theta$ is the angle that the photon forms with respect to the incoming
beams in the centre-of-mass frame then we define $\xt$ and the rapidity $y$ as
\beq
x_T = \frac{2 \,E_T}{\sqrt S}, \quad y=-\log\tan\frac{\theta}{2}, \quad {\rm
where\ }\quad  \et=|{\bf p}| \sin\theta
\eeq
Prompt-photon production takes place both by hard-photon emission from
initial- or final-state quarks (direct component), and by collinear radiation
from final-state partons.  This last mechanism is not fully calculable in
perturbation theory and, in fact, it depends upon the photon fragmentation
function. However, in the cases of practical interest, the corrections due to
the fragmentation processes are small, and we shall limit our considerations
to the hard-photon part. In this approximation,
the differential prompt-photon production cross section integrated
over $y$ is given by the following factorization formula
\beqn
\label{1pxsgamma}
\fl \frac{d\sigma_{\gamma}\(x_T,E_T\)}{d E_T} = \frac{1}{E_T^3} \sum_{a,b}
\int_0^1 dx_1 \;f_{a/H_1}\(x_1,\mu_F^2\)
\,\int_0^1 dx_2 \,\;f_{b/H_2}\(x_2,\mu_F^2\) 
\nonumber \\
\times \int_0^1 dx \,
\delta\!\left(x - \frac{x_T}{{\sqrt {x_1 x_2}}} \right) 
{\hat \sigma}_{ab\to {\gamma}}\(x, \as\(\mu^2\); E_T^2, 
\mu^2, \mu_F^2\) .
\eeqn
where $a,b,c$ denotes the parton indices $(a=q,{\bar q},g)$, and
$f_{a,b/H}\(x,\mu_F^2\)$ are the parton densities of the colliding hadrons,
evaluated at the factorization scale $\mu_F$.  The rescaled partonic cross
sections ${\hat \sigma}_{ab\to \gamma} \equiv E_T^3 \, d{\hat \sigma}_{ab\to
\gamma} / dE_T$ are computable in perturbative QCD  as power series
in the running coupling $\as\(\mu^2\)$, $\mu$ being the
renormalization scale in the $\MSB$ renormalization scheme. Their analytic
expression is known till order $\as^2$~\cite{Aurenche}.

At leading order, only three subprocesses contribute to the direct-photon
production
\beq
\label{loproc}
q + {\bar q} \to g + \gamma \;, 
\quad
q + g \to q + \gamma \;,
\quad
{\bar q} + g \to {\bar q} + \gamma \;.
\eeq
Near the partonic threshold region $x \to 1$, i.e.~when $E_T$ reaches its
maximum value, the partonic cross section are enhanced by double-logarithmic
corrections, due to soft-gluon-radiation suppression
\beq
{\hat \sigma}^{(n)}(x) \sim 
{\hat \sigma}^{(0)}(x) \left[ a_{n,2n} \; \ln^{2n} (1-x) + 
a_{n,2n-1} \ln^{2n-1} (1-x) 
+ \dots \right].
\eeq
Resummation of these soft-gluon effects to all orders in perturbation theory
can be important to improve the reliability of the QCD predictions.

The resummation program of soft-gluon contributions has been carried 
out in the Mellin-transform space, or $N$-space. 
Working in $N$-space, we can disentangle the soft-gluon effects in the parton
densities from those in the partonic cross section and we can
straightforwardly implement and factorize the kinematic constraints of
energy and longitudinal-momentum conservation.  The $N$-moments are defined
as follows
\beq
\label{shn}
\sigma_{\gamma, \,N}(E_T) \equiv \int_0^1 dx_T^2 \;(x_T^2)^{N-1} 
\;E_T^3 \frac{d\sigma_\gamma(x_T,E_T)}{d E_T} \,,
\eeq 
and, in  $N$-space, Eq.~(\ref{1pxsgamma}) takes a simple factorized form
\beq
\label{1pxsngamma}
\fl
\sigma_{\gamma, \,N}\(E_T\) = \sum_{a,b} f_{a/H_1,\,N+1}\(\mu_F^2\)
\;f_{b/H_2,\,N+1}\(\mu_F^2\)   {\hat \sigma}_{ab\to \gamma,
\;N}\(\as\(\mu^2\); E_T^2, \mu^2, \mu_F^2\).
\eeq
The threshold region $x_T \to 1$ corresponds to the limit $N \to \infty$
in $N$-space. In this limit, a hierarchy in $\ln N$ appears and the
partonic cross sections can be rewritten as
\beq
\label{sigreslo}
{\hat \sigma}_{ab\to \gamma, \;N} = 
{\hat \sigma}_{ab\to \gamma, \;N}^{({\rm res})} \left[ 1 + {\cal O} (\as/N)
\right] \;,
\;\;\;ab = q{\bar q},\, qg, \,{\bar q}g \;,
\eeq
where ${\cal O}(\as/N)$ denotes terms that contribute beyond LO and are 
furthermore suppressed by a relative factor ${\cal O}(1/N)$ at large $N$.
The {\em all-order} resummation formulae can be written as
\beq
\label{gammaresqq}
{\hat \sigma}_{ab\to \gamma, \;N}^{({\rm res})} = \alpha \;\as
\;{\hat \sigma}_{ab\to d \gamma, \;N}^{(0)} 
\;C_{ab \to \gamma} \;\Delta_{N+1}^{ab \to d \gamma},
\eeq
where we dropped all the scale dependences, for ease of notation, and we
define 
\beq
\label{deltanll}
\fl
\Delta_N^{ab \to d \gamma} =
\exp \lg \ln N \; g_{ab}^{(1)}\(b_0\,\as\ln N\) +
g_{ab}^{(2)}\(b_0\,\as\ln N\) + {\cal O}\(\as\(\as \ln N\)^k\) \rg \;\;,
\eeq
\beq
\label{cgamma}
\fl
C_{ab\to \gamma} = 
1 + \sum_{n=1}^{+\infty} \; 
\left( \frac{\as}{\pi} \right)^n
C_{ab\to \gamma}^{(n)}
= 1 + \;\frac{\as}{\pi} \;
C_{ab\to \gamma}^{(1)} + {\cal O}\(\as^2\)\,,
\eeq
$b_0$ being the first coefficient of the QCD $\beta$-function.  The function
$\ln N \,g_{ab}^{(1)}$ resums all the {\em leading} logarithmic (LL)
contributions $\as^n \ln^{n+1} N$ in the exponent, $g_{ab}^{(2)}$ contains
the {\em next-to-leading} logarithmic (NLL) terms $\as^n \ln^n N$, and so
forth.  The analytic expressions for the functions $g_{ab}^{(1)}$,
$g_{ab}^{(2)}$ and $C_{ab\to \gamma}^{(1)}$ have been computed in
Ref.~\cite{CMN98}.

\section{Hadron-level results}
\label{HLResults}
\begin{figure}
\begin{center}
  \centerline{ \epsfig{file=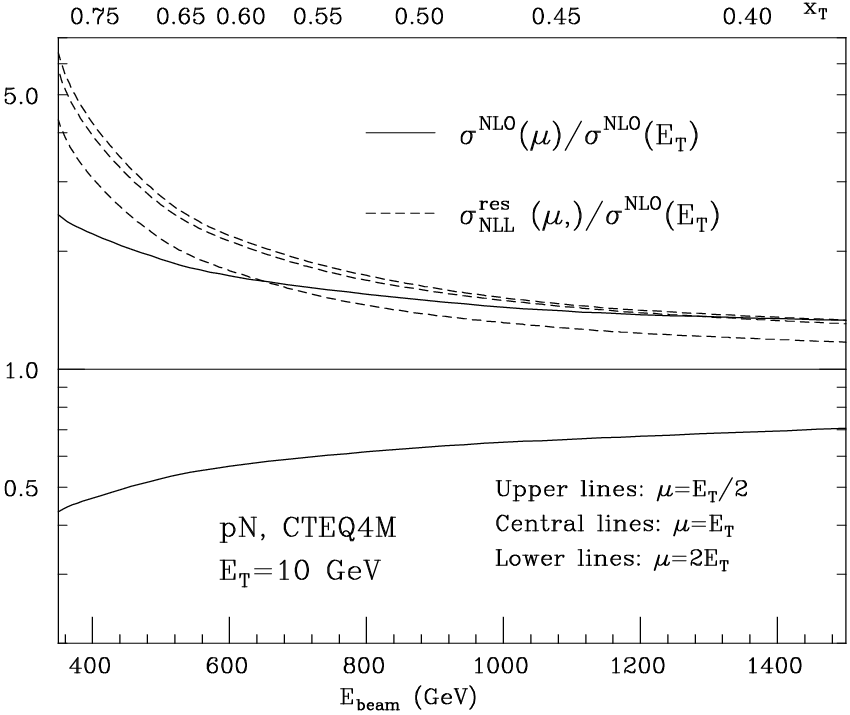,width=0.5\textwidth,clip=}
    \epsfig{file=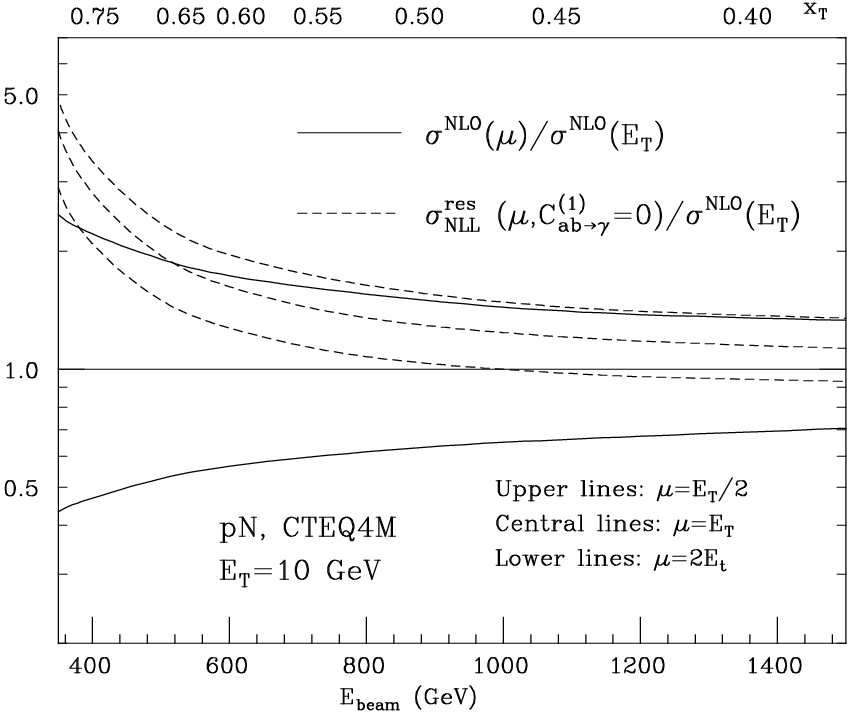,width=0.5\textwidth,clip=} }
  \vspace*{\mysep} 
\caption{\label{fig:scdeph0} Relative size and scale dependence of
$d\sigma/dE_T$ ($qg+q\bar{q}$ components) for prompt photons in $pN$
collisions, at $E_T=10$~GeV, plotted as a function of the proton-beam energy,
$E_{\rm beam}$ (the associated values of \xt\ are given on the top scale).
The solid lines represent the exact NLO result for different choices of
$\mu=\mu_F$ ($\mu=\et/2$ and $2\et$), normalized to the $\mu=\et$ result.
The dashed lines represent the NLO+NLL result (in the right panel we have set
$C_{ab\to\gamma}^{(1)}=0$) for different choices of $\mu$, normalized to the
NLO $\mu=\et$ result.}
\end{center}                                  
\begin{center}
\epsfig{file=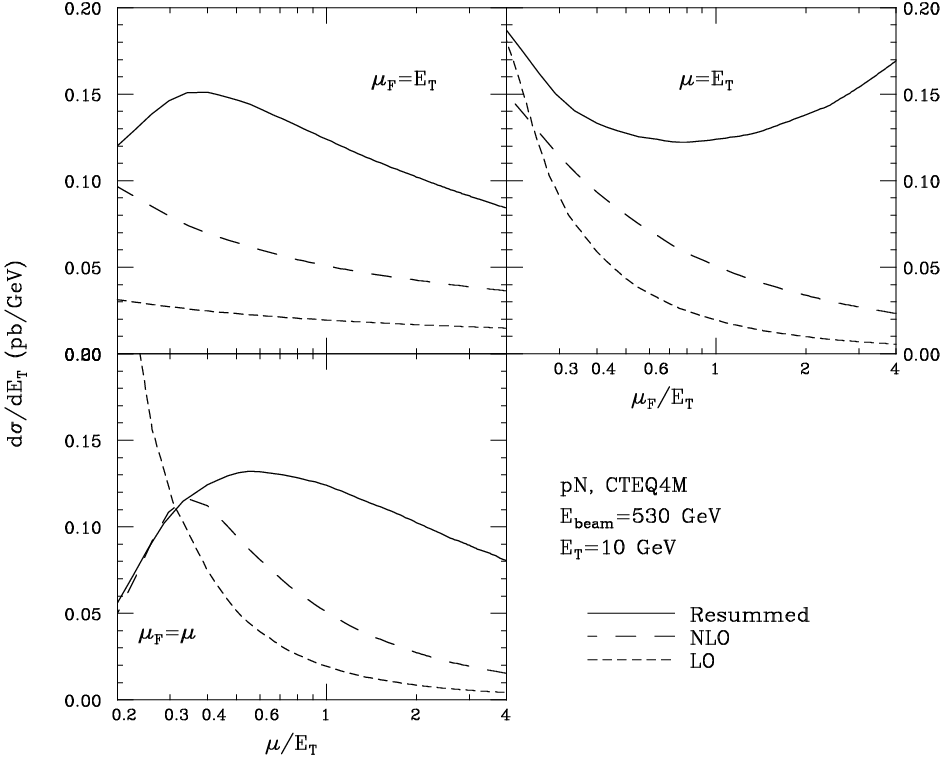,width=0.8\textwidth,clip=}
\vspace*{\mysep}      
\caption{\label{fig:scdep10} 
Scale dependence of the differential $E_T$ distribution in $pN$
collisions, for $E_T=10$~GeV. 
We compare the results at the Born and NLO level with the results
of the resummed calculation.
Upper left: 
renormalization-scale dependence, 
with the factorization scale fixed to 
$\muf=E_T$. Upper right: 
factorization-scale dependence, 
with the renormalization scale fixed to 
$\mu=E_T$. Lower left:
scale dependence, with the renormalization and factorization scales equal.}
\end{center}                                  
\end{figure}

The main points we intend to highlight for the full hadronic cross sections
are: 
\begin{enumerate}
\item the size of the NLL corrections, relative to the NLO contributions;
\item the scale dependence at NLL order.
\end{enumerate}
The size and scale dependence of the resummed cross section ($qg+q\bar{q}$
contributions), compared to the NLO one, is given in Fig.~\ref{fig:scdeph0}.
We plot the distributions as a function of the beam energy
($E_{\rm beam}$) for the fixed value of $\et=10$~GeV. Different values of
$E_{\rm beam}$, therefore, probe different ranges of \xt, as indicated by the
upper labels on the plots.
Note the significant reduction in scale dependence in both cases: with and
without the inclusion of the constant term $C_{ab\to \gamma}^{(1)}$.

We also explored the renormalization- and factorization-scale
dependence of our calculations.  The results, for $pN$ collisions at $E_{\rm
beam}=10$~GeV, are shown in Fig.~\ref{fig:scdep10}. A significant improvement
in the stability of the results, relative to the dependence at LO and NLO, is
observed in all cases.

\ack 
The work presented here has been done in collaboration with S.~Catani,
M.~L.~Mangano, P.~Nason and W.~Vogelsang.

\end{document}